\documentclass[sigconf, 10pt]{acmart} 

\renewcommand\footnotetextcopyrightpermission[1]{}

\usepackage{balance}
\usepackage{amsmath}
\usepackage{url}
\usepackage{makecell}
\usepackage{scalerel}
\usepackage{balance}
\usepackage{epstopdf}
\usepackage{graphicx}
\usepackage{subfigure}
\usepackage{color}
\usepackage{tabularx}
\usepackage{hyperref}
\usepackage{ragged2e}  
\newcolumntype{Y}{>{\RaggedRight\arraybackslash}X}
\usepackage{booktabs}
\usepackage{algorithm}
\usepackage[noend]{algpseudocode}
\usepackage{fixltx2e}
\usepackage{paralist}
\usepackage{listings}
\usepackage{multirow}
\usepackage{enumerate}
\usepackage{bbm}
\usepackage{xspace}
\usepackage{comment}
\usepackage[inline]{aplcomments}
\usepackage{caption} 

\usepackage{comment}
\usepackage[labelfont=bf]{caption}
\MakeRobust{\Call}
\usepackage{enumitem}
\usepackage{etoolbox} 
\usepackage{mdwlist} 

\newcommand{\val}[1]{{\small \texttt{#1}}}

\newcommand{\sj}{\textsc{Auto-Tag}\xspace}

\DeclareMathOperator*{\avg}{avg}

\newcounter{definition}
\newenvironment{definition}[1][]{\refstepcounter{definition}\par\smallskip\textsc{Definition~\thedefinition.\ #1}}{\smallskip}

\newcounter{example}
\newenvironment{example}[1][]{\refstepcounter{example}\par\smallskip\textsc{Example~\theexample.\ #1}}{\smallskip}

\newcounter{theorem}

\newcounter{proposition}

\newtoggle{fullversion}
\toggletrue{fullversion}

\makeatletter
  \newcommand\figcaption{\def\@captype{figure}\caption}
  \newcommand\tabcaption{\def\@captype{table}\caption}
\makeatother

\begin{document}

\iftoggle{fullversion}{}{
\setlength{\floatsep}{0pt}
\setlength{\textfloatsep}{0pt}
\setlength{\abovecaptionskip}{0pt}
\setlength{\abovedisplayskip}{0pt}
\setlength{\belowdisplayskip}{0pt}
\setlength{\itemsep}{0pt}
\setlength{\partopsep}{0pt}
}

\pagenumbering{gobble}

\title{Auto-Tag: Tagging-Data-By-Example  in Data Lakes}

\author{Yeye He, Jie Song, Yue Wang, Surajit Chaudhuri \\ Vishal Anil, Blake Lassiter, Yaron Goland, Gaurav Malhotra}
\affiliation{\institution{Microsoft Corporation}}

\begin{abstract}
As data lakes become increasingly popular in large enterprises today,
there is a growing need to tag or classify data 
assets (e.g., files and databases) in data lakes with additional
metadata (e.g., semantic column-types), as the inferred metadata can
enable a range of downstream applications like data
governance (e.g., GDPR compliance), and dataset search.
Given the sheer size of today's enterprise data lakes 
with petabytes of data and millions of data assets, it
is imperative that data assets can be ``auto-tagged'', using
lightweight inference algorithms and minimal user input.

In this work, we develop \sj{}, a corpus-driven approach
that automates data-tagging of \textit{custom} data types 
in enterprise data lakes.
Using \sj{}, users only need to provide \textit{one} example column
to demonstrate the desired data-type to tag.
Leveraging an index structure
built offline using 
a lightweight scan of the data lake,
which is analogous to pre-training in machine learning,
\sj{} can infer suitable data
patterns  to best ``describe'' the underlying ``domain'' of the given
column at an interactive speed, 
which can then be used to tag additional data of the same ``type'' 
in data lakes.
The \sj{} approach can adapt to custom data-types, and is 
shown to be both accurate and efficient. Part of \sj{} ships
as a ``custom-classification'' feature in a 
cloud-based data governance and catalog 
solution \textit{Azure Purview}.
\end{abstract}

\maketitle

\vspace{-2mm}
\section{Introduction}

Large enterprise data lakes are increasingly common today,
often with petabytes of data and
millions of data assets (e.g., flat files or databases).
Given their sheer sizes,
it has become increasingly important
to \textit{govern} and \textit{catalog} data lakes,
as evidenced by a growing number of offerings 
from startups and established vendors,
such as Azure Purview~\cite{purview}, 
AWS Glue Catalog~\cite{AWSGlue}, Google Cloud Data Catalog~\cite{GoogleCatalog}, 
Alation~\cite{Alation}, Waterline~\cite{Waterline}, 
Collibra~\cite{Collibra}, etc.

A key challenge in governing data lakes is data \textit{tagging} 
(also known as \textit{classification}),  which is the process of
inferring rich metadata (e.g. semantic column-types) from data.
Such inferred metadata are critical for downstream
applications such as data governance
and data discovery:

\begin{figure}
        \centering
        \includegraphics[width=0.4\textwidth]{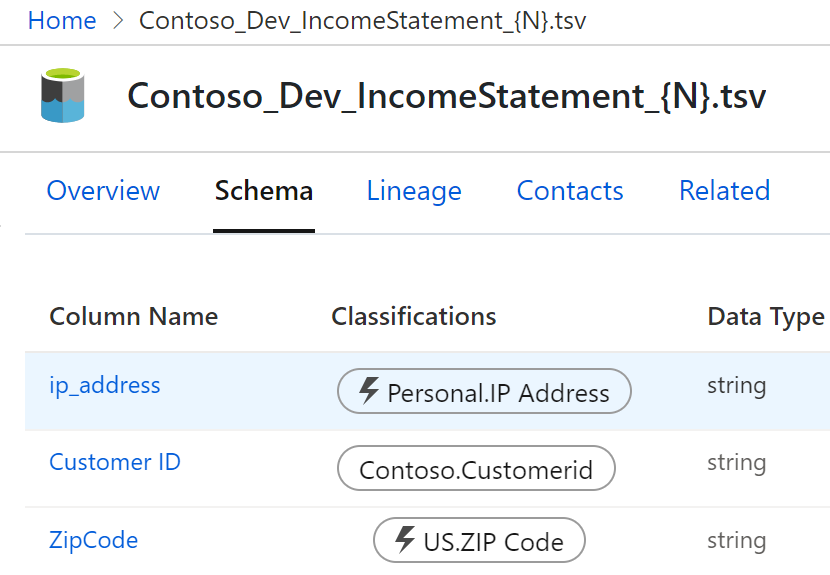}
\caption{Azure Purview: sample columns in an example file ``Contoso\_Dev\_IncomeStatement.tsv'', are automatically 
tagged as ``Personal.IP Address'', 
``Contoso.CustomerId'', ``US.ZIP Code'', etc.}
\label{fig:adc-tagging}
\end{figure}

\underline{\textit{Data governance.}} Data protection regulations
such as GDPR, PCI and CCPA impose strict requirements 
on how sensitive personal data can be
retained and accessed. 
To ensure compliance, it is imperative that enterprises can 
automatically identify sensitive data assets in their data lakes, 
so that these data assets can be governed
in accordance with regulatory requirements.

\underline{\textit{Data discovery.}}  In order to improve the
productivity of enterprise workers, it is increasingly important
for enterprise workers to discover and leverage data assets relevant
to their tasks, using self-service mechanisms
such as data-set search. Given the large number of
data assets in modern enterprise data lakes, and their nondescript, 
sometimes cryptic, nature, datasets search is clearly
challenging (compared to the web search for exampled)~\cite{halevy2016managing}. Suitable metadata
tags/classifications associated with data assets can significantly
improve search relevance, and enhance the overall usefulness of 
enterprise data lakes.

\begin{figure*}[t]
        \centering
        \includegraphics[width=1\textwidth]{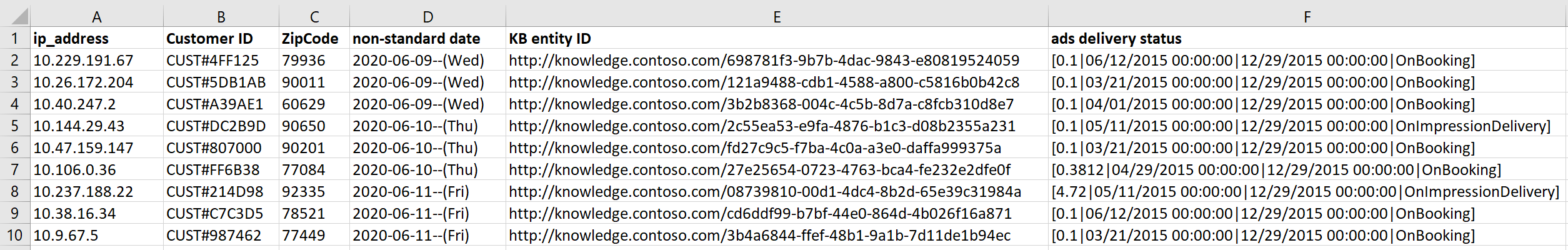}
\caption{An example spreadsheet from Contoso, with many enterprise-specific ``custom'' data-types.}
\label{fig:demo-data}
\end{figure*}

\begin{figure}
        \centering
        \includegraphics[width=0.5\textwidth]{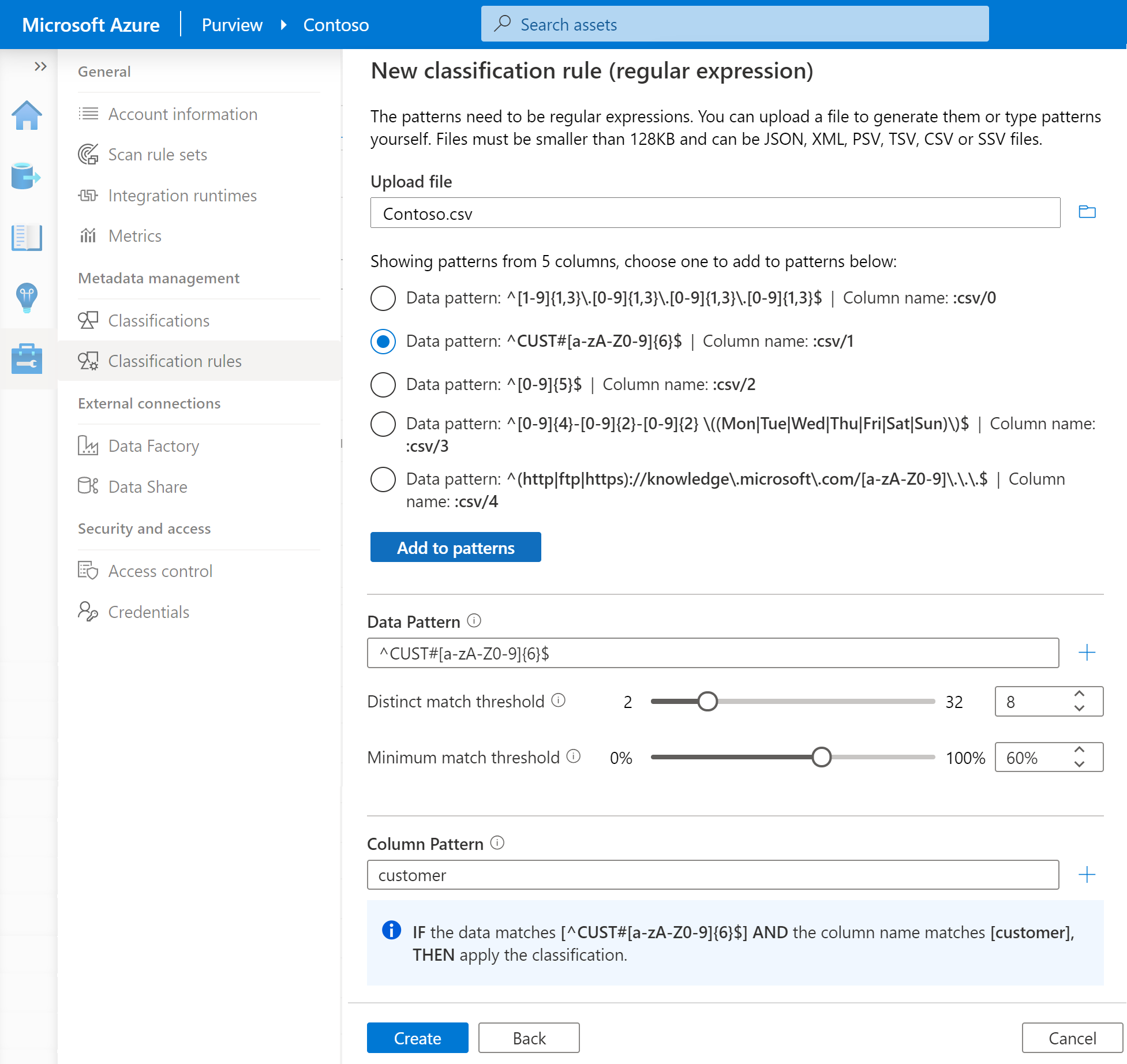}
\caption{Azure Purview UI for custom-classification by-example: after uploading an example data file, columns with inferred patterns
will be suggested for user to inspect and approve.}
\label{fig:purview-demo}
\end{figure}

\textbf{Auto-tagging of ``standard'' data-types.}
Given the importance of data tagging, it is no surprise that leading
vendors in this space all have features
relating to automated data-tagging. 

Figure~\ref{fig:adc-tagging} shows a data-tagging feature
in Azure Purview~\cite{purview}. Out of the box, the system can
already recognize 100+ \textit{standard}
data types commonly found in the public domain~\cite{purview-builtin-types}.\footnote{Note that many of these
data-types are sensitive in nature, making them particularly
relevant to data governance and catalog vendors.}
In this particular example, it detects a few sample 
columns to be of type 
``IP Address'', ``US Zip Code'', etc.
Note that these are well-known data-types from
the public domain, henceforth refer to as
``\textit{standard}'' data-types.

Because standard data-types are well-known and
their corresponding ``data-taggers''
can be reliably tested beforehand, auto-tagging
features for standard data-types are
readily available and work well ``out-of-the-box''
in today's data-catalog and governance
vendors (using a combination
of techniques like predefined regex patterns, bloom-filters, etc.).

\begin{figure*}[t]
        \centering
        \includegraphics[width=1\textwidth]{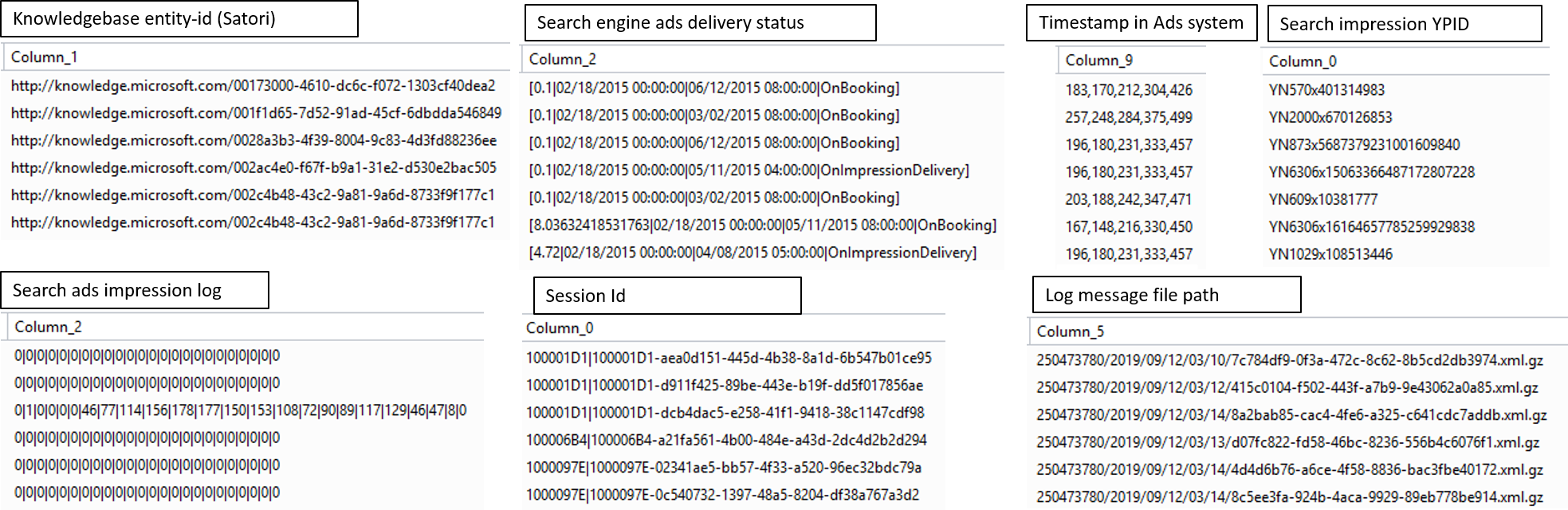}
\caption{Example ``custom'' data types, crawled 
from an enterprise data-lake at Microsoft. Each
column has a distinctive data pattern in proprietary
formats, encoding specific meanings. 
These custom data types are all common, 
occurring in at least 5000 columns in our sample crawl with 7M columns.}
\label{fig:domain-example}
\end{figure*}

\textbf{Auto-tagging of ``custom'' data-types.}
While standard data-types
capture an important class of use cases,
we observe that there is also a
large number of ``\textit{custom}'' enterprise data types,
which are unique to different industries and enterprises.

In the example of Figure~\ref{fig:demo-data}, column B contains unique IDs that this company Contoso (a fictional company) assigns to their customers -- in this case, these values have a prefix of ``\val{CUST\#}'' followed by six hexadecimal characters (e.g., ``\val{CUST\#0FF125}''). Ideally, we want to tag columns of this type as a new custom data-type ``\val{Contoso.CustomerId}'', like shown in the second column of Figure~\ref{fig:adc-tagging}. However, Contoso's way of identifying customers is likely unique, as other companies may devise their own unique-identifiers for customers -- for example, another company may use customer-identifiers that have a string prefix of ``\val{C-}'' 
followed by a unique nine-digit number (e.g., ``\val{C-123456789}''), while yet another company may choose to use other types of UUID. These different forms of customer-ids are custom-made (generated by some programs) and unique to each enterprise, which are thus not well-known ``standard'' data-types that a general-purpose data-catalog solution can possibly anticipate.

There are a large number
of such custom data-types in today's enterprises.
Figure~\ref{fig:domain-example} shows a few
example custom data-types, harvested from
a real production data lake at Microsoft~\cite{chaiken2008scope}. 
Each column here has a distinctive data pattern, which uniquely identifies
a custom data-type widely used inside the 
company. For instance, the first column
is known as Knowledge-Base 
entity-id (Satori~\cite{gao2018building}), which is a unique
ID assigned to real-world entities and used by the search
engine Bing. Similarly, the second column encodes the 
delivery status of Bing ads, etc.

Methods developed for tagging ``standard'' data types
are clearly inapplicable for these 
idiosyncratic ``custom'' 
data types, because they are unique to different enterprises, and
unlikely to be found from the public domain or other enterprises.

This motivates us to look into ways that can tag
custom enterprise data-types, with minimal input from users.

\textbf{Tagging custom-types by-examples.}
Unlike auto-tagging of standard
data types, which can be expected to work 
out-of-box, we believe that tagging custom enterprise
data types requires some amount of human
input, (e.g. from data-owners or domain-experts), to (1) determine
relevant data of interest to ``tag'', 
and (2) provide suitable and meaningful 
tags that can describe the underlying meaning
of the custom data-type (i.e., an algorithm may infer 
values of the form ``\val{CID-12345}''
to be a unique data-type, but cannot be sure of its meaning).

We believe that a human-in-the-loop approach to tagging
custom data-types have two key desiderata:

$\bullet$ \textit{Low human-cost}. The system should 
require minimal input from enterprise users, ideally needing
users to provide only one example column to demonstrate
the custom data type of interest (e.g., an example column 
in Figure~\ref{fig:domain-example}). Note that this is 
different from typical machine-learning tasks -- asking users
to repeatedly provide feedback in the form 
of positive/negative labels to tag \textit{one} custom data type
can be too costly in this setting.

$\bullet$ \textit{Low execution-cost}. It is also important that
any tag-inference algorithm needs to be 
lightweight, in order for the feature to be cost-effective
on large enterprise data lakes. 
Although a deep analysis (e.g., a full scan) 
of the data lake will typically 
yield better predictive accuracy, the scale of the data 
(e.g., petabytes) makes a full scan too expensive.
Thus, auto-tagging algorithms should only perform a \textit{lightweight}
scan (e.g. a row-wise sample per asset), 
in order for the feature to be viable in terms of COGS.

In Section~\ref{sec:related}, we will discuss why existing techniques (e.g.,~\cite{cortez2015annotating, yan2018synthesizing, zhao2019auto, hulsebos2019sherlock, zhang2019sato})
may be insufficient in such a setting, either due to 
high human-costs, or high execution-cost.

In this work, we develop an initial version of this auto-tagging feature
called \sj. Unlike standard machine-learning,
\sj{} has the advantage of requiring only \textit{one} 
labeled example column (low human costs), 
and unlike content-based or dictionary-based approaches, 
does not require a full scan of data files (low execution costs), 
because patterns can be reliably generalized from small samples.

Using a variant of an algorithm we develop in~\cite{auto-validate-full},
we first perform lightweight (row-wise sampled) scans of data lakes
offline, to build a succinct index structure that is analogous to
pre-training in machine-learning. At online inference time,
given an example column of interest users point us to, 
\sj{} leverages the offline index to produce relevant data patterns
that can accurately describe the underlying domain
of the custom type of interest.

Figure~\ref{fig:purview-demo} shows a 
by-example auto-tagging feature in Azure Purview, 
which uses this technology to auto-tag custom data-types. 
From the UI, users can easily upload a data
file with target columns of interest. Leveraging a
succinct index structure built offline, a list of 
suggested data-tagging patterns can be produced
at an interactive speed, so that users can pick the
desired pattern corresponding to the column of interest,
inspect the suggested pattern, before
approving the data-tagging rule. The tagging logic
so created would then be
used to tag additional columns in the data lake 
matching the given pattern during data scans.

Our experiments using real data from a production enterprise data lake
at Microsoft~\cite{chaiken2008scope} suggest that
\sj{} is both accurate and cost-efficient, for tagging custom data-types.
We report experimental results in Section~\ref{sec:experiments}.

\vspace{-3mm}
\section{Related Works}
\label{sec:related}
Auto-tagging of data assets is
an important topic, given the abundance of data in data lakes today. 
We will review related data-tagging techniques 
below\footnote{We focus on techniques that produce 
tags based on \textit{data-values in columns}.
Orthogonal techniques leveraging other types of information
also exist, e.g., 
program-flows~\cite{sen2014bootstrapping}.}, and
discuss why they are not immediately
applicable to our problem (high human-cost
or execution cost).

We emphasize that different classes of techniques 
below are often suitable for
orthogonal types of data (e.g., natural-language content
vs. machine-generated data), and thus do not subsume each other.

\textbf{Data-tagging by value-patterns.}
It is reported that a substantial fraction of enterprise 
data columns have regex-like
patterns~\cite{auto-validate-full}, for which 
pattern-based approaches are the most suitable.

There are many existing techniques from the \textit{data profiling}
literature, which infers patterns based on example data-values.
These include research prototypes like 
Potter's wheel~\cite{raman2001potter}, 
X-System~\cite{ilyas2018extracting},
LearnPads~\cite{fisher2005pads, fisher2008dirt},
FlashProfile~\cite{padhi2018flashprofile},
and commercial implementations
like Microsoft SQL 
Server SSIS~\cite{ssis-profiling}, 
Trifacta~\cite{trifacta-profiling}, 
Ataccama~\cite{ataccama}.

As we will highlight in Section~\ref{sec:auto-tag},
the goal of data-profiling is distinctively different from data-tagging --
it aims to find patterns 
to succinctly summarize \textit{given data values only},
which tend to produce overly-specific (or
under-generalized) patterns, which yield
low recall when used for auto-tagging.
There is significant room for improvement,
and is the focus of our corpus-driven \sj{} approach.

\textbf{Data-tagging by machine learning models.} 
Machine-learning or deep-learning based approaches,
such as Auto-EM~\cite{zhao2019auto}, 
Sherlock~\cite{hulsebos2019sherlock} and 
Sato~\cite{zhang2019sato},
have been developed to tag
columns with natural-language
content (e.g., company-names, people-names, etc.). Such approaches, however, are often a poor fit for
machine-generated data (e.g., GUID, employee-ID, etc.), 
and would complement pattern-based approaches.
Such approaches also typically require a non-trivial amount of
labeled data, increasing the cost of adoption for tagging custom enterprise
data-types (high human costs).

\textbf{Data-tagging by value-overlap.} 
Techniques have also been developed to tag columns
based on value overlap in enterprise 
tables~\cite{cortez2015annotating}
and web tables~\cite{venetis2011recovering, taneva2013mining, wang2012understanding},
where the idea is that if a substantial fraction of values in a given column 
match a known dictionary of values (e.g., a known list
of department-names or product-names), then the column can 
be tagged accordingly. 

When such ``dictionaries of values'' for enterprise
concepts are not known a priori,
techniques are developed to harvest such ``dictionaries'' for data-tagging.
These techniques are known in the literature as
\textit{set expansion}~\cite{he2011seisa, pantel2009web, wang2007language}, \textit{concept discovery}~\cite{li2017discovering, ota2020data}, and more broadly
knowledge-base 
construction~\cite{hoffart2013yago2, bollacker2008freebase, dong2014knowledge, gao2018building}.
These approaches, however, 
typically require full-scans for high recall, 
thus introducing high execution-costs.

\textbf{Data-tagging by synthesized programs.} 
Because values of certain 
data types (e.g., credit-card numbers, UPC codes) 
have unique signatures such as check-sums,
which can only be detected via specific program-logic,
program-synthesis based data-tagging have been proposed,
which synthesize type-detection functions using 
existing code~\cite{yan2018synthesizing}.
Such approaches, however, requires the presence 
of a enterprise-specific code repository to be effective.

\nocite{zheng2013learning, madhavan2008google, zhao2015smartcrawler, he2013crawling, naumann2002, ritze2015, sen04, shepherd2012sando, Simmhan05, singhal2012, suchanek2007yago, talaika2015, zheng2006, Hazewinkel01, gartner, Informatica, Talend, trifactatypes, li2020deep, mudgal2018deep, konda2018magellan}

\vspace{-3mm}
\section{Auto-Tag By-Examples}
\label{sec:auto-tag}

Given the need of low human-costs
and execution-costs discussed above,
in this work we set out to solve the auto-tagging problem,
for string-valued custom-types with syntactic patterns
(our prior study~\cite{auto-validate-full}
suggests that this is an important class 
accounting for around 40\% string-valued columns
in a production data lake).

We will first briefly describe the
pattern language used.

\subsection{Preliminary: Pattern Language}
\label{sec:language}
We use a standard pattern language (similar 
to~\cite{raman2001potter}). Other languages 
can also be plugged in \sj{} to produce corresponding
patterns.

Figure~\ref{fig:hierarchy} shows a standard generalization 
hierarchy, where leaf-nodes represent the English alphabet, 
and intermediate nodes (e.g., 
\val{<digit>}, \val{<letter>}) represent \textit{token} that values
can generalize into. A pattern is a sequence 
of (leaf or intermediate) tokens,
and for a given value $v$, this hierarchy
induces a space of all patterns consistent with $v$, 
denoted by $\mathbf{P}(v)$.
For instance, for a value $v =$ ``\val{9:07}'', we could generate
$\mathbf{P}(v) = $ \{
 ``\val{<digit>:<digit>\{2\}}'', 
 ``\val{<digit>+:<digit>\{2\}}'', 
 ``\val{<digit>:<digit>+}'', 
 ``\val{<num>:<digit>+}'',  ``\val{9:<digit>\{2\}}'', \ldots
\}, among many other options.

Given a column $C$ for which patterns need
to be generated, we define the 
space of \textit{candidate patterns}, 
denoted by $\mathbf{P}(C)$, as the set of patterns 
consistent with values  $v \in C$. 
We use an in-house implementation to produce patterns based
on the hierarchy in Figure~\ref{fig:hierarchy} (other hierarchies 
and languages can be applied similarly in \sj{}).


\vspace{-2mm}
\subsection{Find Suitable ``Domain'' Patterns}
\label{sec:intuition}

Given the pattern language $\mathbf{P}$ described above, and 
given a data lake, consisting of a large collection of tables $\mathbf{T}$
(which can be flat files such as .csv, .tsv, .xls, .json, as well as
database files and database tables, etc.), at a high level
our auto-tagging problem can be stated as follows.

\begin{definition}
Auto-tag by-examples. Given a data lake of tables $\mathbf{T}$,
users demonstrate a desired action to tag data of type $t$,
by providing one example
column $C \in \mathbf{T}$ consisting of a set of 
values $C=\{v_i\}$ that are of type $t$,  and a tag 
$n(t)$ describing this type $t$.
Let $\mathbf{P}(C)$ be the set of all data-patterns consistent
with $C$, our goal is to select a suitable pattern
$p \in \mathbf{P}(C)$, such that for any data column 
$D \in \mathbf{T}$, if $p \in \mathbf{P}(D)$ or $p$ also 
matches the column $D$,
then $D$ is also likely of type $t$ (can be tagged 
as $n(t)$).
\end{definition}

\begin{figure}
        \centering
        \includegraphics[width=0.4\textwidth]{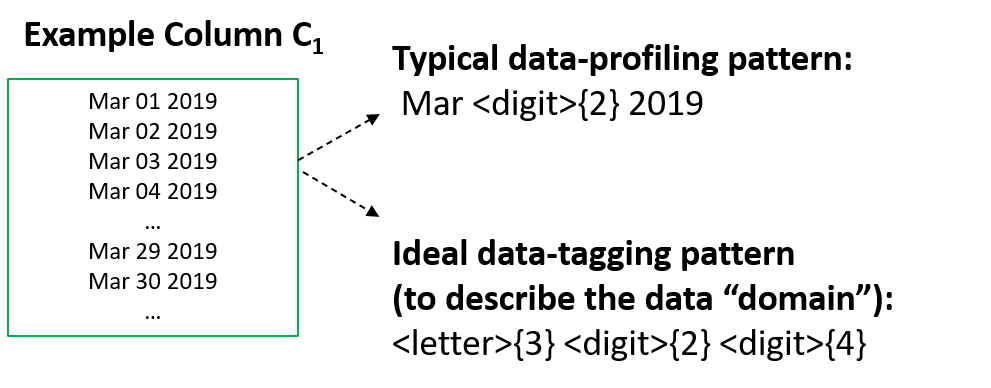}
\caption{Example showing different patterns produced for data-profiling vs. data-tagging, because the two have very different objectives.}
\label{fig:datetime-example}
\end{figure}

\begin{example}
As a concrete example, users provide an 
example column $C_1$ shown in Figure~\ref{fig:datetime-example},
as well as a tag $n(C_1)$, say ``\textit{date}''.
The system should suggest a suitable pattern
``{\small \texttt{<letter>\{3\} <digit>\{2\} <digit>\{4\}}}''
that best describe the underlying data ``domain''.
Once this is reviewed and approved by users, it can be used
to tag additional columns in $\mathbf{T}$ matching the same pattern
(with tag ``\textit{date}'').
\end{example}

\begin{figure}
\vspace{-6mm}
        \centering
        \includegraphics[width=0.35\textwidth]{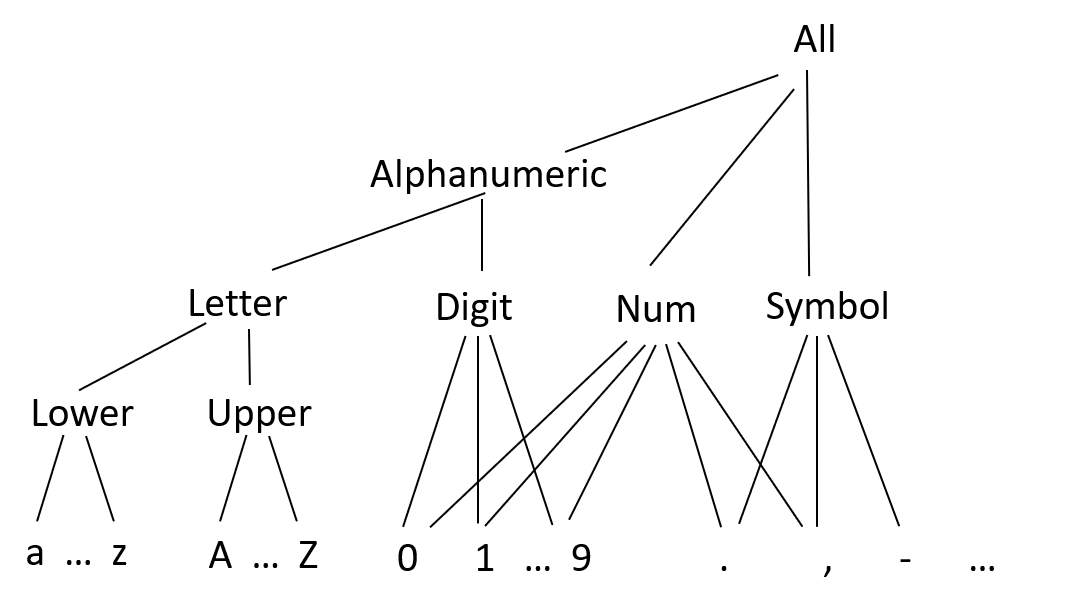}
\caption{Example generalization hierarchy.}
\label{fig:hierarchy}
\end{figure}

\begin{figure}
        \centering
        \includegraphics[width=0.5\textwidth]{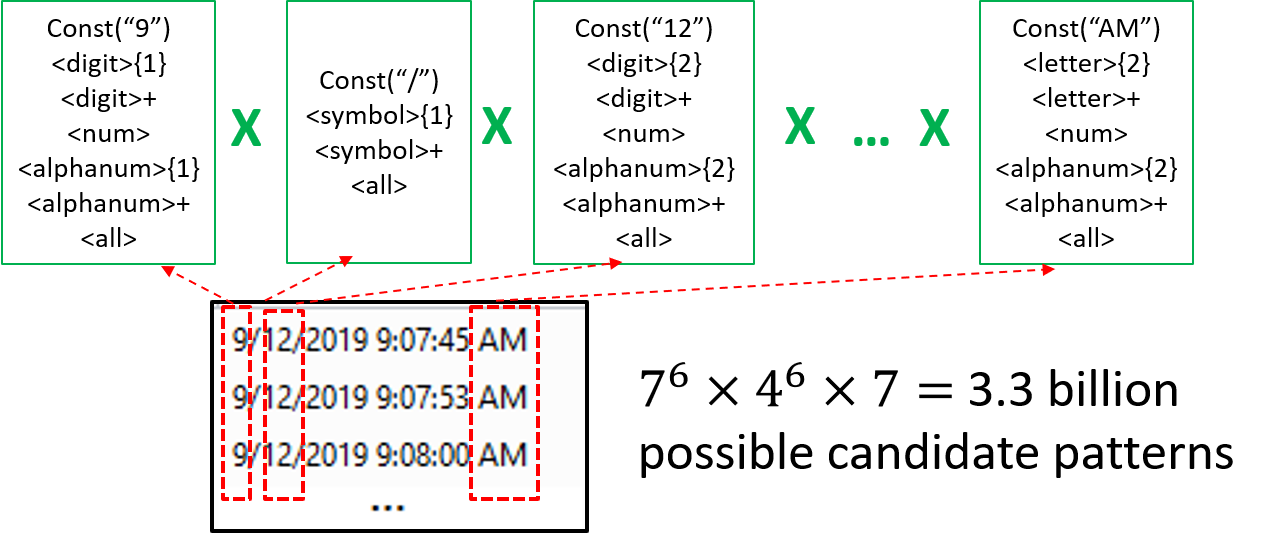}
\caption{Possible ways to generalize a column of date-time strings, using the
hierarchy in Figure~\ref{fig:hierarchy}.}
\vspace{-3mm}
\label{fig:generalization}
\end{figure}

One challenge is that $\mathbf{P}(C)$ is  large
(there are many ways to ``generalize'' a column $C$
into patterns). For a simple column of date-time strings
like in Figure~\ref{fig:generalization}, and using a standard generalization 
hierarchy as in Figure~\ref{fig:hierarchy}, one could produce over 3 billion
possible patterns. For example, the first part
(digit ``\val{9}'' for month) alone can be generalized in 7 different
ways shown in the top-right box of Figure~\ref{fig:generalization},
and the cross-product at each position
creates a large space (3.3 billion patterns) for this
seemingly simple column.

Given a large space of candidates $\mathbf{P}(C)$, the key is to:

(1) Not ``\textit{under-generalize}'': or use
overly restrictive patterns, which lead to low recall for data-tagging; and

(2) Not ``\textit{over-generalize}'': or use overly generic patterns
(e.g. the trivial ``\texttt{.*}''), which lead to low precision.

These are the key reasons why related techniques 
like \textit{pattern-profiling} 
(e.g., Potter's Wheel~\cite{raman2001potter},
PADS~\cite{fisher2005pads}, X-System~\cite{ilyas2018extracting}, 
FlashProfile~\cite{padhi2017flashprofile}, etc.) are not directly
applicable to data-tagging, because they have
very different objectives.

Specifically, 
the goal of pattern-profiling is to succinctly ``summarize''
a given set of values in column $C$, so that users can quickly understand
what is in $C$ without needing to scroll/inspect the entire $C$.
Such techniques explicitly consider \textit{only} values in $C$,
\textit{without needing to consider values not present in $C$}
(e.g., other valid values that are in the same ``domain'' as $C$).

For example, classical pattern profiling
methods like Potter's Wheel~\cite{raman2001potter}
and FlashProfile~\cite{padhi2017flashprofile} would correctly
generate a desirable pattern 
``{\small \texttt{Mar <digit>\{2\} 2019}}'' for $C_1$ in 
Figure~\ref{fig:datetime-example}, which is valuable
from a pattern-profiling's perspective as it succinctly describes all given
values in  $C_1$.
However, this pattern is not suitable for 
data-tagging, as it under-generalizes 
and would miss many similar date-time columns
like ``{\small \texttt{Apr 01 2019}}'', thus yielding 
low recall. A more appropriate
data-tagging pattern should instead describe the entire
``domain'' of possible values for this data-type, e.g.,
``{\small \texttt{<letter>\{3\} <digit>\{2\} <digit>\{4\}}}''.

A key challenge here is to select suitable
patterns from $\mathbf{P}(C)$, when only one
example column $C$ is given. This is intuitively
difficult if we only look at $C$ -- for the date examples in 
Figure~\ref{fig:datetime-example}, 
we as humans know the ideally-generalized pattern for this type, 
but for data from proprietary domains with ad-hoc formats
(e.g., Figure~\ref{fig:domain-example}), even
humans may find it hard and need to use
additional evidence to reason about ideal patterns to describe 
the corresponding ``domain''
(e.g., by inspecting similar-looking columns in the lake $\mathbf{T}$).

Following this intuition,  we propose a 
corpus-driven approach \sj{} that
leverages summary statistics of
$\mathbf{T}$ (with similar-looking columns) 
to determine the best pattern, which we describe next.

\subsection{Auto-Tag: Estimate Pattern Quality}
Intuitively, a pattern
$p(C) \in \mathbf{P}(C)$ is a good domain
pattern if it captures 
all valid values from the same domain,
and a ``bad'' pattern if it under-generalizes
or over-generalizes.

\textbf{Avoid under-generalization.}
We show that it is possible to infer whether  $p(C)$
under-generalizes, using summary statistics from
$\mathbf{T}$ (without human input).

\begin{figure*}
        \centering
        \includegraphics[width=0.95\textwidth]{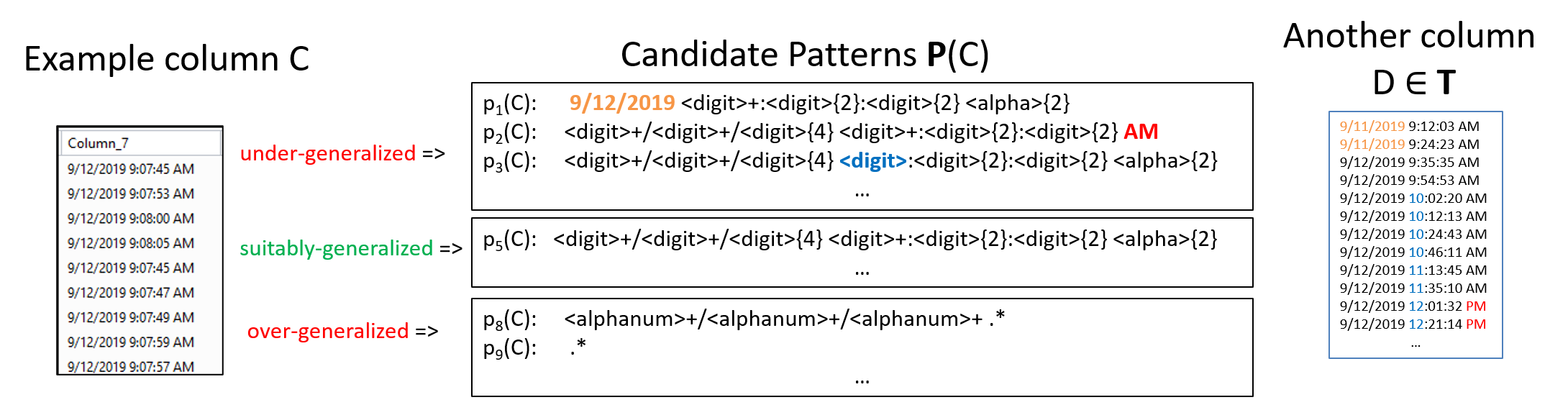}
\caption{Given a column $C$, leverage tables in a data lake $\mathbf{T}$, to infer whether candidate patterns $p(C)$ over-generalize or under-generalize.}
\vspace{-3mm}
\label{fig:intuition}
\end{figure*}

\begin{example}
\label{ex:intuition} 
The left of
Figure~\ref{fig:intuition} shows a query column $C$ for which
domain patterns need to be generated. A few
candidate patterns in $\mathbf{P}(C)$ are listed in
the middle. In this example, we know
that $p_1(C)$, $p_2(C)$, $p_3(C)$ are ``bad'' because
they under-generalize the domain (too ``narrow'').

We show that this can be inferred using $\mathbf{T}$ alone.
Specifically, $p_1(C)$ likely under-generalizes the domain,
because we can find many columns like $D \in \mathbf{T}$ 
shown on the right of Figure~\ref{fig:intuition}
that are ``impure'' -- these columns contain
values that match $p_1(C)$, as well as values that do not
(e.g., ``{\small \texttt{9/11/2019 09:12:03 AM}}'',
where the day part does not match $p_1(C)$).
A large number of ``impure'' columns likely
indicate under-generalizations.

We can show that $p_2(C)$ also likely under-generalizes the domain,
as it makes many columns like $D$ ``impure'' 
(the ``{\small \texttt{PM}}'' part does not match $p_2(C)$).

The same can be said about $p_3(C)$
(values like ``\val{10:02:20 AM}'' are inconsistent
with $p_3(C)$ because they have two-digit hours, whereas
$p_3(C)$ uses a single \val{<digit>}).

Using $p_5(C)$ to describe the domain, 
on the other hand, would not yield many
``impure'' columns in $\mathbf{T}$, suggesting that it
does not under-generalize the domain.
\end{example}

Intuitively, we can use the \textit{impurity} of $p$ on
data columns $D \in \mathbf{T}$,
measured as the fraction 
of values in $D$ not matching $p$,
to infer whether $p$ is an under-generalization:
\begin{definition}
\label{def:impurity}
The \textit{impurity} of a candidate pattern $p$ on 
a data column $D \in \mathbf{T}$, 
is defined as:
\begin{equation}
\label{eqn:impurity}
\text{Imp}_{D}(p) = \frac{|\{v | v \in D, p \notin \mathbf{P}(v)\}|}{|\{ v | v \in D \}|}
\end{equation}
\end{definition}

\begin{example}
\label{ex:fnr}
In Figure~\ref{fig:intuition}, 
$\text{Imp}_D(p_1)$ can be calculated as $\frac{2}{12}$,
since the first 2 values (with ``\val{9/11/2019}'') 
out of 12 do not match $p_1$.
Similarly, $\text{Imp}_D(p_3)$ can be calculated as $\frac{8}{12}$,
since the last 8 values in $D$ (with two-digit hours) 
do not match $p_3$.

Finally, $\text{Imp}_D(p_5)$ is $\frac{0}{12}$,
since all values in $D$ match $p_5$.
\end{example}

We note that if $p(C)$ is used to tag
data in the same domain as $C$, then 
Imp$_D(p)$ directly corresponds to
expected false-negative-rate (FNR), or recall-loss
for data-tagging tasks.

\begin{definition}
\label{def:fnr}
The expected \textit{false-negative-rate} (FNR) of using pattern $p(C)$ 
to tag a data column $D$ drawn from the same domain as $C$,
denoted by $\text{FNR}_{D}(p)$, is defined as:
\begin{equation}
\label{eqn:fnr}
\text{FNR}_{D}(p) = \frac{\text{FN}_{D}(p)}{\text{TP}_{D}(p)+\text{FN}_{D}(p)}
\end{equation}
Where TP$_{D}(p)$ and  FN$_{D}(p)$  are the number of 
false-positive detection and true-negative detection of $p$ on $D$,
respectively. 
\end{definition}
Note that since $D$ is from the same domain as $C$,
ensuring that TP$_{D}(p)$ and  FN$_{D}(p)$ = $|D|$, 
$\text{FNR}_{D}(p)$ can be rewritten as:
\begin{equation}
\label{eqn:fnr}
\text{FNR}_{D}(p) = \frac{|\{v | v \in D, p \notin \mathbf{P}(v)\}|}{|\{v | v \in D \}|} = \text{Imp}_D(p)
\end{equation}
Thus allowing us to estimate $\text{FNR}_{D}(p)$ using $\text{Imp}_D(p)$.

\begin{example}
\label{ex:fnr-impurity}
Continue with Example~\ref{ex:fnr}, it can be
verified that the expected 
FNR of using $p$ as the domain pattern for $D$,
directly corresponds to the impurity $\text{Imp}_D(p)$ --
e.g., using $p_1$ to tag $D$ has $\text{FNR}_{D}(p_1)$
= $\text{Imp}_D(p_1)$ = $\frac{2}{12}$; while using 
$p_5$ to tag $D$ has $\text{FNR}_{D}(p_5) = \text{Imp}_D(p_5) = 0$, etc.
\end{example}

Based on FNR$_{D}(p)$ defined for one column $D \in \mathbf{T}$, 
we can in turn define the estimated FNR on the entire corpus $\mathbf{T}$,
using all column $D \in \mathbf{T}$ 
where some value $v \in D$ matches $p$:
\begin{definition}
\label{def:fnr_corpus}
Given a corpus $\mathbf{T}$, we estimate the 
FNR of pattern $p$ on  $\mathbf{T}$, 
denoted by FNR$_{\mathbf{T}}(p)$, as:
\begin{equation}
\label{eqn:fnr_agg}
\text{FNR}_{\mathbf{T}}(p) = \avg_{D \in \mathbf{T}, v \in D, p \in \mathbf{P}(v) }{\text{FNR}_{D}(p)}
\end{equation} 
\end{definition}


\begin{example}
\label{ex:fnr_t}
Continue with the $p_5$ in Example~\ref{ex:fnr}
and Example~\ref{ex:fnr-impurity}. Suppose 
there are 5000 data columns $D \in \mathbf{T}$ 
where some value $v \in D$ matches $p_5$.
Suppose 2000 columns out of the 5000 have $\text{FNR}_{D}(p_5) = 0$, 
and the remaining 3000 columns have $\text{FNR}_{D}(p_5) = 50\%$. 
The overall $\text{FNR}_{\mathbf{T}}(p_5)$ 
can be calculated as $\frac{3000 * 50\%}{5000} = 30\%$,
using Equation~\eqref{eqn:fnr_agg}.
\end{example}

\textbf{Avoid over-generalization.}
So far we have focused on avoiding under-generalization.
Similarly we should also avoid over-generalization, 
such as $p_8$ and $p_9$ shown in 
Figure~\ref{fig:datetime-example}. We achieve
this by measuring \textit{coverage} of pattern $p$
over $\mathbf{T}$.

\begin{definition}
\label{def:coverage}
The \textit{coverage} of a candidate pattern $p$ on $\mathbf{T}$, 
is defined as:
\begin{equation}
\label{eqn:coverage}
\text{Cov}(p) = |\{D | D\in \mathbf{T}, p \in \mathbf{P}(D)\}|
\end{equation}
\end{definition}

Intuitively, among all candidate patterns that do not under-generalize
(using impurity-based estimates), we should pick
the pattern with the least coverage, which is the least likely to over-generalize.

\begin{example}
\label{ex:coverage}
Recall that in Example~\ref{ex:intuition}, we
infer that $p_1(C)$, $p_2(C)$ and $p_3(C)$ likely under-generalize
(thus can be excluded), 
while $p_5(C)$, $p_8(C)$ and $p_9(C)$ do not. For the remaining
patterns, given a data lake
with 10M columns, we find the coverage of 
$p_5(C)$, $p_8(C)$ and $p_9(C)$ to be 20K, 500K and 10M, respectively.
We can then pick $p_5(C)$ as the suitable pattern for auto-tagging,
as it does not under-generalize, and at the same time is the 
least likely to over-generalize.
\end{example}

Given this intuition, we formalize pattern-inference
as an optimization problem below.

\vspace{-2mm}
\subsection{Problem Formulation: CMDT}
We now describe the basic version 
of \sj{} as follows. 
Given an input query column $C$ and a background corpus $\mathbf{T}$,
we need to produce a domain pattern $p(C)$, 
such that $p(C)$ is expected to have a low FNR
but also with few false positives.
We formulate this as an optimization problem, 
called Coverage-Minimizing 
version of Data-Tagging (CMDT), defined as:
\begin{align}
\hspace{-1cm} \text{(CMDT)} \qquad{} \min_{p \in \mathbf{P}(C)} & \text{Cov}_\mathbf{T}(p)   \label{eqn:cmdt_obj} \\ 
 \mbox{s.t.} ~~ & \text{FNR}_\mathbf{T}(p) \leq r \label{eqn:cmdt_fnr} \\
 & \text{Cov}_\mathbf{T}(p) \geq m \label{eqn:cmdt_cov}
\end{align}

Equation~\eqref{eqn:cmdt_fnr} states that the expected
recall loss of using $p$ as the domain pattern for $C$, 
estimated from $\text{FNR}_\mathbf{T}(p)$, 
is lower than a given threshold $r$. 
Equation~\eqref{eqn:cmdt_cov} is an optional constraint
that requires the coverage of $p$, 
$\text{Cov}_\mathbf{T}(p)$, defined as 
the number of columns in $\mathbf{T}$ that match $p$,
to be greater than a given threshold 
$m$ (otherwise the custom data-type may be too
niche to be interesting).

The domain pattern $p$ we produce for $C$ is
then the minimizer
of CMDT from the space of all candidate patterns
$\mathbf{P}(C)$ (Equation~\eqref{eqn:cmdt_obj}),
which as discussed minimizes the chance of over-generalization
(and false-positives in auto-tagging) for a given recall
constraint.

We should note that the CMDT formulation is 
closely related to the FMDV problem in~\cite{auto-validate-full}.
The two problems share the same problem structure but use 
different objective functions (tailored to data-tagging
and data-validation, respectively).
We leverage similar vertical-cut and
horizontal-cut algorithms in~\cite{auto-validate-full}, 
and also optimization methods (lightweight scan with 
offline indexing). Together, these mechanisms 
achieve (1) interactive response time and (2) cost effectiveness
(by scanning a small fraction of rows per file).
We refer 
readers to~\cite{auto-validate-full} for details of the 
algorithms in the interest of space.

\vspace{-2mm}
\section{Experiments}
\label{sec:experiments}
We implement our offline indexing algorithm in a Map-Reduce-like language
called Scope~\cite{zhou2012scope} with UDFs in C\#, executed on a 
production cluster~\cite{chaiken2008scope}. 

\vspace{-2mm}
\subsection{Benchmark Evaluation}
\textbf{Data set}. We evaluate algorithms using 
a real corpus $\mathbf{T}$ with 
7.2M data columns, crawled from a production data
lake at Microsoft~\cite{zhou2012scope}.

\textbf{Evaluation methodology}.  
We randomly sample 1000 columns 
from $\mathbf{T}$
to produce a benchmark set of columns, denoted by
$\mathbf{B}$. We use 
the first 1000 values of each column to control 
column size variations.

Given a benchmark $\mathbf{B}$ with 1000 columns, 
$\mathbf{B} = \{C_i | i \in [1, 1000]\}$,
we manually assign a ground-truth tag-id for each column.
This produces clusters of columns in the same data-type
and should be assigned same tags.

We then evaluate precision and
recall of patterns generated on $\mathbf{B}$ 
as follows. For each column $C_i \in \mathbf{B}$, 
we use the first $10\%$ of values in $C_i$ (or 100 values)
as the ``training data'',
denoted by $C_i^\text{train}$, from which 
patterns need to be generated. 
Each algorithm $A$ can observe $C_i^\text{train}$ 
and ``learn'' pattern $A(C_i^\text{train})$.
The inferred pattern is denoted as $A(C_i^\text{train})$.

To test recall of $A(C_i^\text{train})$ when
$C_i$ is used for auto-tagging, denoted by $R_A(C_i)$, we use the remaining
$90\%$ of values from $C_i$, as well as other columns
in $\mathbf{B}$ with the same ground-truth cluster-id.
These are data columns drawn from the same data-type as $C_i$,
which we expect $A(C_i^\text{train})$ to match.
We take chunks of 100 values from these columns
as column-units, and compute recall by testing the fraction of
column-units that can be correctly tagged
(under different matching thresholds).

To test precision, denoted by $P_A(C_i)$, we use columns in 
$\mathbf{B}$ with a different ground-truth cluster-id,
which are from a different data-type as $C_i$.
We know that it is a false-positive 
if $A(C_i^\text{train})$ were to tag these columns.
We compute precision accordingly, by taking 
chunks of 100 values from these columns as column-units,
and compute the fraction of
column-units that not incorrectly tagged by $A(C_i^\text{train})$.

The overall precision/recall on benchmark $\mathbf{B}$ 
is the average across all cases:
$P_{A}(\mathbf{B})  = \avg_{C_i \in \mathbf{B}} P_A(C_i)$, and
$R_{A}(\mathbf{B})  = \avg_{C_i \in \mathbf{B}} R_A(C_i)$.
Both of these are between 0 and 1 as usual.

\vspace{-2mm}
\subsection{Methods Compared}
\label{sec:methodsCompared}
We compare the following algorithms using
benchmark $\mathbf{B}$, by reporting precision/recall numbers
$P_A(\mathbf{B})$ and $R_{A}(\mathbf{B})$.

\textbf{\sj}. This is our proposed approach using CMDT.

\textbf{Potter's Wheel (PWheel)}~\cite{raman2001potter}. This is
an influential pattern-profiling method that finds the best
pattern based on 
minimal description length (MDL). 

\textbf{SQL Server Patterns}~\cite{ssis-profiling}. SQL Server 
has a data-profiling feature in SSIS and Data Tools. 
We invoke it programmatically
to produce regex patterns
for each column.

\textbf{XSystem}~\cite{ilyas2018extracting}. This recent
approach develops a flexible branch and
merge strategy to pattern profiling. We use the authors' implementation
on GitHub~\cite{xsystem-code} to produce patterns.

\textbf{FlashProfile}~\cite{padhi2018flashprofile}. FlashProfile
is a recent approach to pattern profiling, which clusters
similar values by a distance score. 
We use the authors' implementation
in NuGet~\cite{FlashProfileCode}.

We also compare with \textbf{Grok}~\cite{grok}, which is a popular
approach that uses a collection
of curated regex patterns 
to detect common types in log messages;
\textbf{schema-matching}~\cite{do2002comparison} based
methods (followed by pattern-profiling);
and a simple \textbf{Value-Union}~\cite{cortez2015annotating} method
that is more suitable for natural-language content. 
These methods are
not as effective (e.g., producing overly-generic patterns
with low precision), and are omitted from the results
(to ensure we can zoom in on the competitive methods in the figures).

\vspace{-2mm}
\subsection{Experimental Results}
We evaluate different methods based on tagging quality
(precision/recall), latency, and memory footprint.

\begin{figure}
        \centering
        \includegraphics[width=0.48\textwidth]{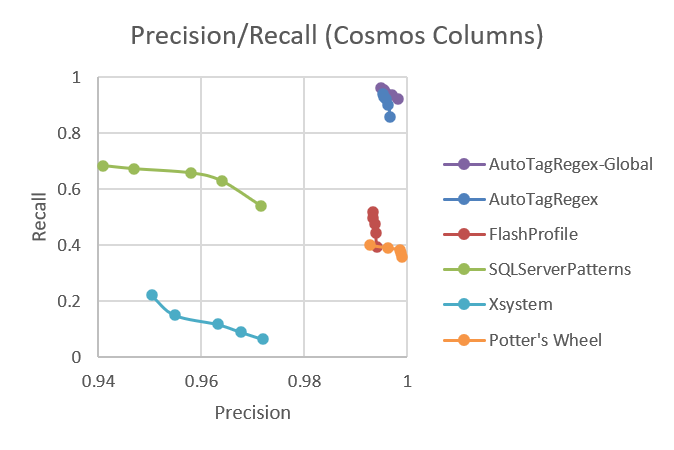}
\caption{Precision/Recall on 1000 randomly
sampled cosmos data. Results are scaled
to 
test columns that have patterns.}
\label{fig:pr}
\end{figure}

\textbf{Quality.}
Figure~\ref{fig:pr} shows precision/recall of all 
methods using the enterprise benchmark $\mathbf{B}$ with 
1000 randomly sampled test cases.
It can be seen that \sj{} is substantially better than other
methods in both precision and recall.

Among all the baselines, we find data profiling techniques like
\texttt{PWheel} and \texttt{FlashProfile} to also be of
high-precision. However, these techniques tend to
under-generalize and produce lower recall
(because as discussed, data profiling techniques
aim to optimize for a fundamentally different objective compared
to data-tagging).

\textbf{Latency.} Given that it is important to 
produce regex suggestions at interactive speed
(for users to inspect and verify), we compare the
mean and max latency of different methods on 
1000 benchmark test columns. It can be seen 
that \sj{} is clearly interactive, where the
max latency is 0.663 second. 

In comparison, methods like
\texttt{FlashProfile} and \texttt{XSystem} use expensive
clustering, which on average take 6-7 seconds
per input column, 
where the max latency per column is close to 6 minutes for both methods.
Note that for both 
\texttt{FlashProfile} and \texttt{XSystem}
we use authors' original implementations~\cite{xsystem-code, FlashProfileCode}.

\textbf{Memory footprint.}
We also evaluate the average and max 
memory usage for pattern-learning
per input column. \sj{} avoids expensive bottom-up
enumeration and is lightweight, which uses an average of 1.9MB memory 
(2.8MB max). In comparison, clustering-based 
pattern-profiling methods like
\texttt{FlashProfile} takes 162MB memory on average,
with a max memory usage of 7.9GB.

\begin{table}[h!]
\vspace{-2mm}
\centering
\small
\begin{tabular}{|c | c | c |} 
 \hline
  Method &  mean-latency (ms) & max-latency (ms) \\  \hline
  \sj  &  \textbf{12}  & \textbf{663}  \\  \hline 
 FlashProfile &  7076 & 359382  \\  \hline
 XSystem &  6411 & 346996  \\  \hline
\end{tabular}
\caption{Mean/max latency on 1000 benchmark cases.}
\vspace{-2mm}
\label{tab:compare-eval}
\end{table}

\vspace{-3mm}
\section{Conclusions}

Observing the need to data-tagging for custom 
data types in enterprise data lakes, we propose a 
corpus-driven \sj approach to
infer relevant data patterns. This is shown to be
accurate and cost-effective, when evaluated 
on real enterprise data from a production data lake.

\clearpage



\begin{small}
\balance
\bibliographystyle{abbrv}
{\footnotesize
\bibliography{auto-tag}
}
\end{small}


\end{document}